\documentclass[graybox]{svmult}

\usepackage{subfig}

\usepackage{amssymb}
\usepackage{amsmath}
\usepackage{array}
\usepackage{mdwmath}
\usepackage{mdwtab}
\usepackage{eqparbox}
\usepackage{graphicx}
\usepackage{caption}
\usepackage[ruled]{algorithm2e} 
\usepackage{epstopdf}
\usepackage{algcompatible}
\usepackage[mathscr]{eucal}
\usepackage[section]{placeins}

\usepackage{type1cm}        
\usepackage{makeidx}         
\usepackage{graphicx}        
\usepackage{multicol}        
\usepackage[bottom]{footmisc}

\usepackage{newtxtext}       %
\usepackage{newtxmath}       

\makeindex             

\begin{document}

\title*{Analysis of Movement-based Connectivity Restoration Problem in Wireless Ad-Hoc and Sensor Networks}
\author{Umut Can Cabuk, Vahid Khalilpour Akram  and Orhan Dagdeviren}
\institute{Umut Can Cabuk \at Ege University, Izmir, Turkey \email{umut.can.cabuk@ege.edu.tr}
\and
 Vahid Khalilpour Akram \at Ege University, Izmir, Turkey \email{vahid.akram@ege.edu.tr}
\and Orhan Dagdeviren \at Ege University, Izmir, Turkey \email{orhan.dagdeviren@ege.edu.tr}
}
%
%
\maketitle

\begin{abstract}\\
Topology control, including topology construction and maintenance phases, is a vital conception for wireless ad-hoc networks of any kind, expressly the wireless sensor networks (WSN). Topology maintenance, the latter phase, concerns several problems, such as optimizing the energy consumption, increasing the data rate, making clusters, and sustaining the connectivity. A disconnected network, among other strategies, can efficiently be connected again using a Movement-based Connectivity Restoration (MCR) method, where a commensurate number of nodes move (or are moved) to the desired positions. However, finding an optimal route for the nodes to be moved can be a formidable problem. As a matter of fact, this paper presents details regarding a direct proof of the NP-Completeness of the MCR Problem by a reduction of the well-studied Steiner Tree Problem using the minimum number of Steiner points and the bounded edge length.
\end{abstract}


\section{Introduction}

As an integral part of the revolutionary Internet of Things (IoT) concept, wireless ad-hoc networks are one of several technologies those are shaping the century. Sensors, actuators, and other lightweight electronic devices with wireless networking capabilities (also called nodes) are used to build smart systems to be used in industries, commercial services and everyday life [1]. However, due to the cost-oriented restrictions of these devices, such as limited battery lifetime, scarce bandwidth, weak processing power, low memory and possibly others, make these nodes prone to various problems, which may degrade the network's stability and even can completely interrupt the services they provide.

As a result of these restrictions; say, a battery discharge, software faults, a hacking/hijacking attack or a physical damage may be occurred, and these may cause removal, displacement or death of the subject node(s). Otherwise, external factors like strong wind, rain, floods, ocean currents, stray/wild animals or curious/malevolent people may cause the same on a larger scale. 

Removal, displacement or scatter of the nodes of a wireless ad-hoc network, may eventually disconnect the whole network, whenever ``enough'' number of nodes have been placed on places far ``enough''. The definition of ``enough'' is totally scenario dependent. Also, death of the nodes may or may not cause a disconnection, depending on the initial placement. The connectivity can be restored essentially in two ways; either new nodes (in sufficient numbers) should be placed in appropriate places to bind (or bridge) the existing nodes, or the existing nodes should be move towards each other and/or to a determined rendezvous-point. 

Both strategies may work well, again depending on the scenario. Besides, both are easy to implement using na\"{i}ve methods, as long as it is possible to add new nodes to the network (for the first solution) or it is possible to move the nodes (for the second solution). But as stated earlier, in such applications, resources are scarce. So, the number of nodes to be added should be minimum (for the first solution), or likewise, the total distance to be travelled by the nodes should be minimum. Please note that the total distance moved is actually a function of the energy consumed and the time spent, which are also scarce resources as well as the node count. In fact, the former strategy is a fundamental variation of the well-known Steiner Tree Problem, and already proven as an NP-Complete [2] problem. The latter (as we call the Movement-based Connectivity Restoration Problem - MCR), however, was studied less and no assertive claims have been found in the literature regarding its complexity class, which is then proven to be also NP-Complete within this work.

The remaining parts of this paper have been organized as follows; In Section II we provide a brief survey about the existing works on the connectivity restoration in wireless ad-hoc and drone networks. The network model and some preliminaries are outlined in Section III. In Section IV, we presented a proof of NP-Hardness of movement-based connectivity restoration problem. Lastly, our conclusions were presented in Section V.

\section{ Related Works}

Connectivity restoration is one the important challenges in wireless ad-hoc networks and many researches have focused on this problem from different perspectives. One of the approaches for connectivity restoration is deploying new nodes [3-14]. In these approaches the aim is restoring a network's connectivity by placing the lowest number of nodes to optimum locations. These approaches are generally used in static networks where the required locations are always reachable by humans. Some other studies pay attention to determine the radio power of nodes to create a connected network [15-18]. The radio hardware limitation is main restriction for these methods because if the distance between disconnected partitions is long, increasing the radio power may be insufficient to restore the connectivity. 

Some other researches focus on the preventing the disconnection of the network by finding current \textit{k} value of a network [19-25].  A network is \textit{k}-connected if it can tolerate at least \textit{k} nodes without losing the connectivity of active nodes. In 1-connected networks losing a single node can separate the network to disconnected partitions. In a 4-connected network at least 4 nodes must stop working to lose the network connectivity. Finding the \textit{k} value and preserving it in a safe level reduces the risk of disconnection however, finding or preserving the \textit{k} value is an energy consuming task which needs a large amount of message passing.  

For mobile ad-hoc networks various movement-based connectivity restoration algorithms have been proposed [26-30] which move mobile nodes to their neighbors or some new locations according different principles,  however, to the best of our knowledge, there is not any optimal algorithm that conducts movement-based connectivity restoration. This process is more complicated for the networks with$\ k>1$. Despite of proposed different heuristics algorithms in different researches for movement-based connectivity restoration,  existing studies do not give a proof of the NP-Hardness nor they present an optimal method. In this study, we present a proof that clearly shows that the movement-based connectivity restoration is NP-Hard for wireless ad-hoc networks .   

\section{ Problem Formulation}

We can model any wireless ad-hoc network as an undirected graph $G=(V,E)$, in which $V$ is a set of vertices (exemplifying wireless sensor nodes) and  $E$ is the set of edges (exemplifying wireless communication links), considering the following assumptions:

\begin{enumerate}
\item  All nodes have access to the medium and have broadcasting capabilities (unicasting is trivial). 
\item  All wireless links are bidirectional.
\item  All nodes have the same communication range \textit{r}.
\item  A link is automatically established between any pair of nodes that have an Euclidean distance of \textit{d} $\mathrm{\le}$ \textit{r}. 
\item  No link can be removed to form a specific topology, unless the above condition is no longer satisfied. In the case of \textit{d} $\mathrm{>}$ \textit{r}, the link is automatically removed.
\item  Operational ranges (i.e. sensing area) are omitted. 
\end{enumerate}

Without loss of generality, the given assumptions make it easier to work on the problem. Nevertheless, the provided proof can easily be expanded considering the absence or change of any of the above statements. A not-to-scale drawing representing the nodes (smaller red circles with numbers inside) and the radio coverage (larger blue circles with dashed lines) of an example wireless ad-hoc network is given in Figure 1.a. The same wireless ad-hoc network is then re-drawn in Figure 1.b using the graph representation, which will be referred in the rest of the paper.

\begin{figure}[!htbp]
\centering
\subfloat[]{\includegraphics[width=0.3\textwidth]{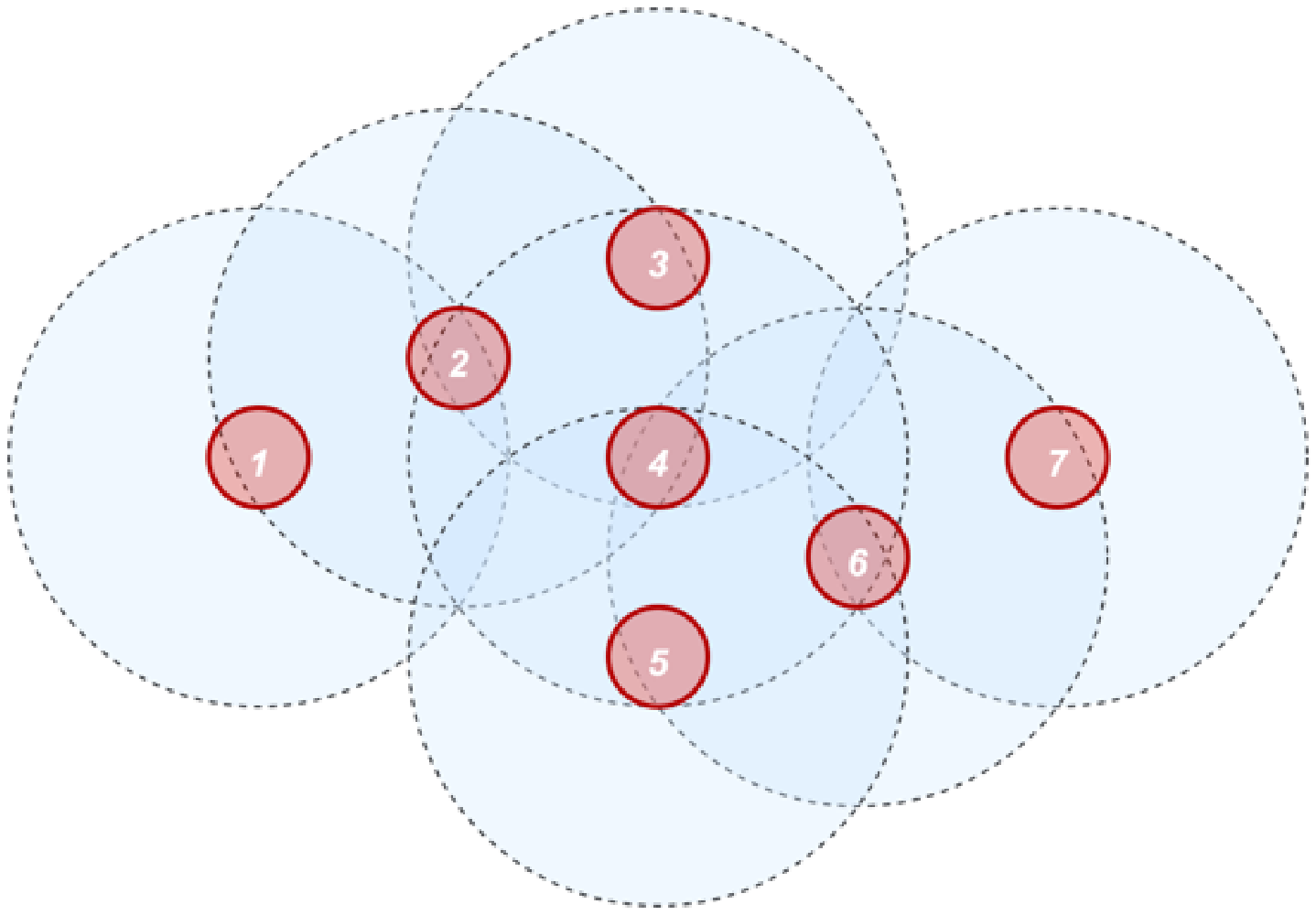}}
\hfil
\subfloat[]{\includegraphics[width=0.3\textwidth]{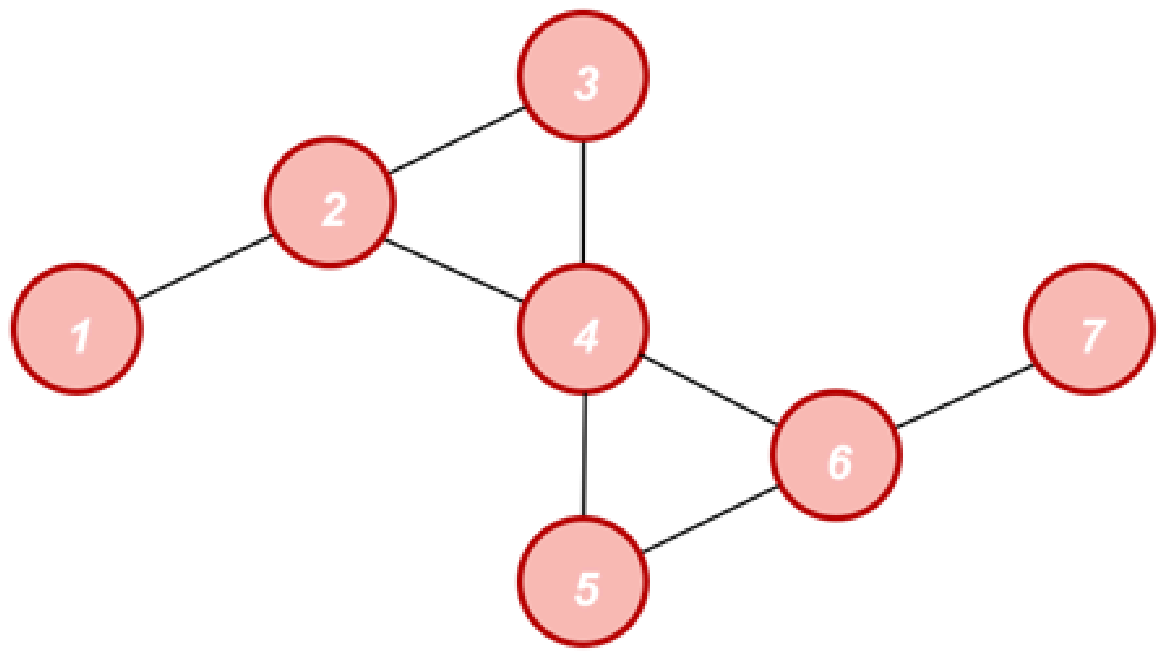}}
\hfil
\caption{a) An example wireless ad-hoc network, showing the nodes and their radio coverage b)Modeling a wireless ad-hoc network as an undirected graph. }
\end{figure}

Table I lists the symbols that are used in this paper, and for the proof. 

\begin{table}[!h]
	\begin{center}
\begin{tabular}{|p{0.4in}|p{1.7in}|} \hline 
\textbf{Symbol} & \textbf{Meaning} \\ \hline 
{\textbar}V{\textbar} & Cardinality of set V \\ \hline 
$U\backslash V$ & Items in U not included in V \\ \hline 
${\mathbb{R}}^2$ & 2-dimensional Euclidean plane \\ \hline 
$\mathbb{R}$ & Real numbers' set \\ \hline 
$\left|\left|v,u\right|\right|$ & Distance btw. points $u$ and $v$ in ${\mathbb{R}}^2$ \\ \hline 
\end{tabular}
\caption{Used Symbols in the Paper}
	\end{center}
\end{table}
\section{NP-Completeness of Movement-based  Connectivity Restoration}

Within this study, we present a proof that clarifies the movement-based connectivity restoration in wireless ad-hoc networks (e.g., sensor networks) is an NP-Complete problem. The Steiner Tree Problem with minimum cardinality of required Steiner points and bounded edge length is already a known NP-Complete problem [2], and we show that it can be reduced to the movement-based connectivity restoration in polynomial time. The Steiner Tree Problem using minimum number of Steiner Points and Bounded Edge Length (ST) can formally be defined as follows:

\textbf{ST Problem: }Let any set $V$ of $n$ points be in ${\mathbb{R}}^2$ and a positive number $r\in \mathbb{N}$, the challenge is constructing a spanning tree $T=(V\cup S,E)$ within the graph \textit{G} with nodes $V$ and the minimum count of Steiner points \textit{S}, so that the maximum length of every edge $e\in E$ is $r$.

Figure 3 below demonstrates an example instance of the ST Problem where $V=\{1,\ 2,\ 3,\ 4,\ 5,\ 6,\ 7\}$ and $E=\emptyset $. In general, \textit{E} does not necessarily be an empty set, as long as \textit{G} is disconnected. The red circles (in both Fig. 3-a and b) represent the initially deployed nodes, while the green circles with dotted circumference (in Fig. 3-b) represent the add-on nodes.
\begin{figure}[!htbp]
\centering
\subfloat[]{\includegraphics[width=0.25\textwidth]{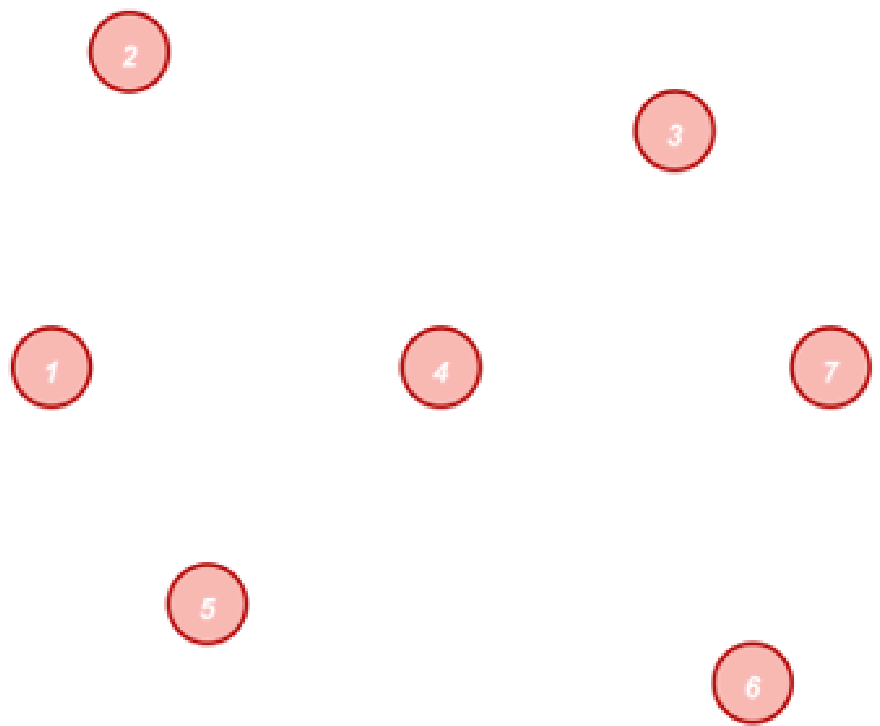}}
\hfil
\subfloat[]{\includegraphics[width=0.25\textwidth]{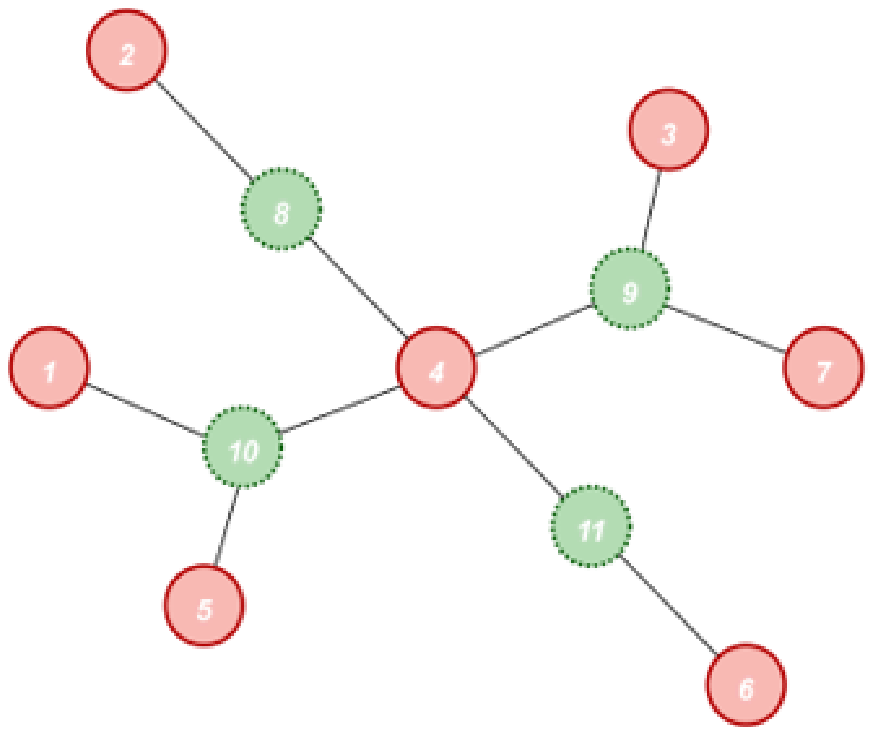}}
\hfil
\caption{a) A disconnected network as an example instance of the ST Problem. b) The same network connected by placing new nodes on arbitrary Steiner points.}
\end{figure}

As given in Figure 3-b; using the pre-calculated Steiner points to place new nodes, provide a connected network using the minimum required number of nodes. Although this solution provides means of cost efficiency in restoring the connectivity; depending on the network type and the scenario, there may be cases where adding new nodes to the network is expensive, or impractical, or not even possible. In these cases, if it is possible to move the nodes or if they can move autonomously in an efficient manner, then addressing the MCR Problem instead can help restoring the connectivity of the network better. 

\textbf{MCR Problem:} Let $V\subset {\mathbb{R}}^2$ be the set of position of nodes of a wireless ad-hoc network, where the communication range of each node is $r$, and $E=\{\left\{u,v\right\}\ \ |\ \ u,v\in V\ \ and\ \left|\left|u,v\right|\right|\le r\}$   be the set of links between the nodes. Given a moving cost function$\ \ c\ :\ V\ \times \ \ {\mathbb{R}}^2\to \mathbb{R}$, the problem is finding a mapping function $M\ :\ V\to {\mathbb{R}}^2$, such that $\sum_{v\in V}{c\left(v,M\left(v\right)\right)}$ is minimum and $G=(M(V),E)$ is connected. 

Figure 4-a shows an example instance of the MCR problem where $V=\{1,\ 2,\ 3,\ 4,\ 5,\ 6,\ 7\}$ and $E=\emptyset $ (again not necessarily). Figure 4-b, 4-c and 4-d give representations of hypothetical solutions to the instance given in Figure 4-a. Even though the drawings are not-to-scale, the edge length condition $\left|\left|u,v\right|\right|\le r$ is exclusively satisfied. A comprehensive run of a valid algorithm optimally (or sub-optimally) solving the MCR problem may find the real results, which include the optimal (or sub-optimal) route and the final position of each node that can move or be moved. There might also be multiple optimal solutions.

\begin{figure}[!htbp]
\centering
\subfloat[]{\includegraphics[width=0.23\textwidth]{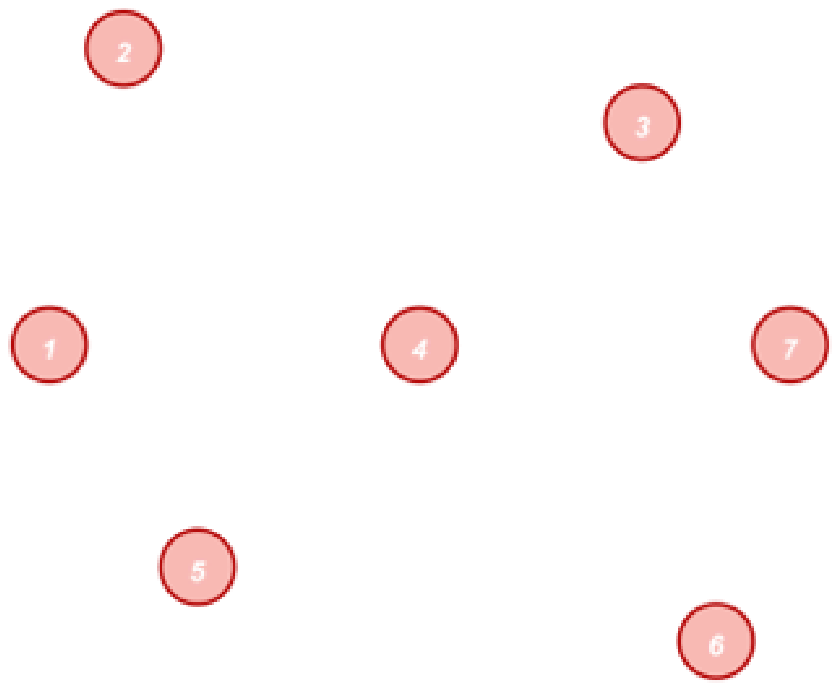}}
\hfil
\subfloat[]{\includegraphics[width=0.23\textwidth]{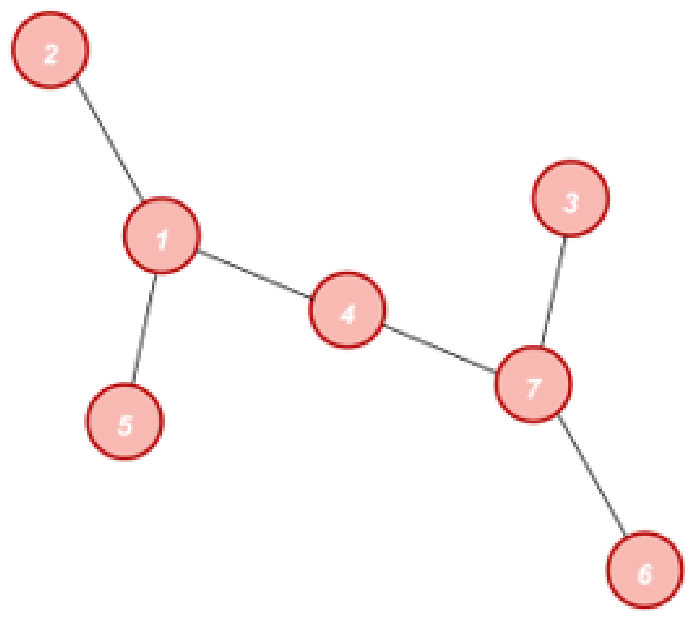}}
\hfil
\subfloat[]{\includegraphics[width=0.23\textwidth]{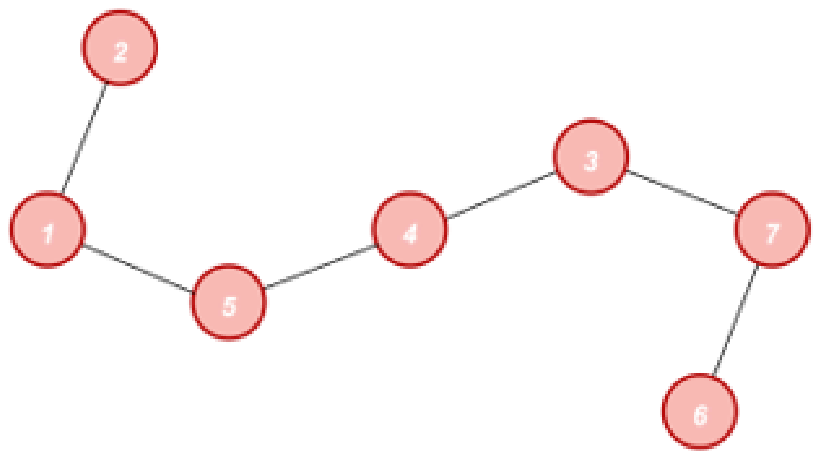}}
\hfil
\subfloat[]{\includegraphics[width=0.23\textwidth]{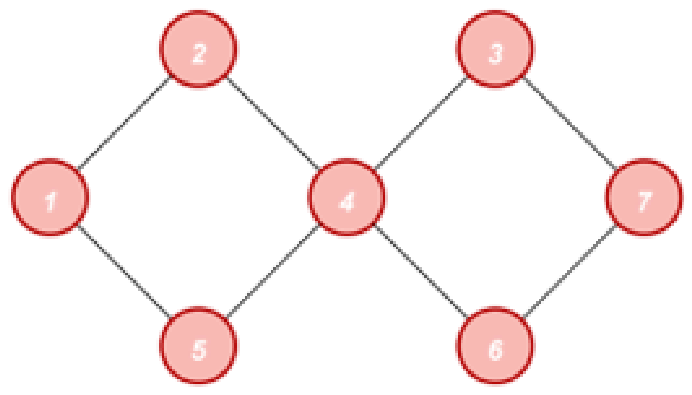}}

\caption{a) A disconnected network as an example instance of the MCR problem. b) A possible connected network in the form of a balanced tree (iff root is 4). c) A possible connected network in the form of a line. d) A possible connected network in the form of a polygon.}
\end{figure}

\textbf{Theorem 1: }\textit{It is possible to construct a polynomial-time reduction from ST Problem to MCR Problem.}

\textbf{Proof:}
Let $V$ be a set of $n$ nodes that has to be contained by a spanning tree utilizing the minimum cardinality of Steiner points and $r$ be the maximum length of any edge in that spanning tree. If it is assumed that the set $\ S$ is an optimal resolution for the given ST Problem, so a hypothetically optimal algorithm for that problem should accurately point out the set $S=\{s_1,\ s_2\ \ ,\dots s_h\}$ of $h\ge 0$ Steiner points, such that $G=(S\cup V,E)$ is connected, where $E=\left\{\left\{u,v\right\}\ \ \right|\ \ u,v\in S\cup V\ \ and\ \left|\left|u,v\right|\right|\le r\}$.  

Assume that we have an optimal algorithm $MCR\ (V,r,c)$ for MCR Problem which returns a set $M$ of new locations of nodes in $V$, such that $\sum_{v\in V}{c\left(v,M\left(v\right)\right)}$ is minimum and $G=(M(V),E)$ is connected. Let $ST(V,r)$ algorithm returns minimum number of Steiner nodes which are required to construct a connected spanning tree. We must show that a polynomial time algorithm can uses $MCR\ (V,r,c)$ to produce optimal solution for $ST(V,r)$. The obtained optimum spanning tree (within the ST Problem) can be classified as a connected graph. The $MCR$ algorithm accepts the initial nodes set, the $r$ value and a cost function $c$. The $V$ and $r$ are common inputs for both $MCR$ and $ST$ algorithms. We define the cost function $c$ as follow: 
\[c\left(v,p\right)=\left\{ \begin{array}{c}
1\ \ \ \ \ \ \ \ \ \ \ \ \ \ \ \ \ \ \ \ \ \ \ \ \ if\ v\ \epsilon \ V\ and\ \ v\neq p \\ 
0\ \ \ \ \ \ \ \ \ \ \ \ \ \ \ \ \ \ \ \ \ \ \ \ \ \ \ \ \ \ \ \ \ \ \ \ \ \ \ \ otherwise \end{array}
\right.\ \ \ \ \]

According the above function the cost of moving node $v\in V$ to a new location $p$ is1. Note that, the cost of moving each node to its desired position (keeping node unmoved) is 0.  The cost of moving any node $v\notin V$ to any point is 0. Using this cost function and MCR algorithm, the following algorithm produces an optimal solution for the ST problem. 

\noindent 

\noindent Algorithm \textbf{ST }($V,\ r$)\textbf{}

1:$U\leftarrow \mathrm{\emptyset }$

2:$M\leftarrow \ \boldsymbol{MCR}(V,r,c)$\textbf{}

3:$\ \boldsymbol{while}\ $ $\sum_{v\in V}{c\left(v,M\left(v\right)\right)}$ $>0\ $\textbf{do}

4:$ \quad U\leftarrow U\ \cup \ t\notin (V\cup U)\ .$ 

5:$ \quad M\leftarrow \ MCR(V\cup U,r,c)$

 6:\textbf{return} $U$.

\noindent 

The above algorithm repeatedly selects a random node$\ t\notin (V\cup U)\ $and adds it to set $U$ and calls $MCR(V\cup U,r,c)$ to find a mapping for \textit{MCR} problem. According to the defined cost function if $MCR$ moves any node $v\in V$  to a new location the resulting total cost will be higher than 0. When the nodes in \textit{U} is not enough for building a connected graph, the \textit{MCR} shall move some nodes in \textit{V} and the total cost will be higher than 0. When \textit{MCR} moves only the newly added nodes and create a connected graph then the resulting total cost will be 0. In other words. when the nodes in $U$ are enough for creating a connected graph, the total cost of moving will be 0.  Formally we can write:

\[\sum_{v\in V}{c\left(v,M\left(v\right)\right)}=0\ \to \ \nexists v\in V\ s.t\ v\neq M\left(v\right)\ \to \ \] 
\[\forall \ u\ \in U\ :\ \ u\in M\ \] 

Since we add the nodes to \textit{U}, one by one, upon the condition $\sum_{v\in V}{c\left(v,M\left(v\right)\right)}=0\ $ becomes true, the set \textit{U} will have the optimal number of Steiner points. The largest iteration count for the "while" loop is $|U\backslash V|$ or the cardinality of appended Steiner nodes. For any$\ n\ge 0$ and $r>0$, an $n\times n$ 2-dimensional Euclidean plane can entirely be covered by employing uttermost ${((n/r)+1)}^2$ circles with radius $r$ such that the spacing between the centers of those neighboring circles becomes at least equal to the radius $r$. So $\left|U\backslash V\right|\le {((n/r)+1)}^2$and the algorithm has polynomial-time complexity.

According to Theorem 1, if we have a polynomial-time algorithm for \textit{MCR} problem, then we can solve the \textit{ST} problem in polynomial time which proves that the \textit{MCR} problem is NP-Hard, because the NP-Hardness of \textit{ST} problem has already been proved.

\section{ Conclusion}
Wireless ad-hoc networks, especially the wireless sensor networks made of constrained devices, are susceptible to displacement and disconnection caused by various internal (i.e. battery outage) and external factors (i.e. storm). The two main approaches used to re-establish the connection are adding new nodes to appropriate positions and moving the existing nodes to appropriate positions (if they are mobile or movable). A hybrid solution may also be possible but neglected in this study.

For the first strategy; yielding an optimal solution that requires the minimum possible number of nodes to reconnect the network is equivalent to the well-studied Steiner Tree Problem and was already proven to be NP-Complete. For the second; finding and optimal solution that requires coverage of the minimum possible distance in total to bring the network together is defined as the MCR Problem. Within the scope of this work; the MCR Problem on a 2-dimensional plane is thoroughly shown to be reducible to the Steiner Tree Problem in polynomial time. Hence, it is proven that the MCR problem is also NP-Complete.

\end{document}